\documentclass[final,5p,times,twocolumn]{elsarticle}

\usepackage{amssymb}
\usepackage{color}
\usepackage{amsmath}
\usepackage{multirow}
\usepackage{booktabs}
\usepackage{bm}
\usepackage{braket}
\usepackage{amsmath}
\usepackage{graphicx}
\usepackage {ulem}
\biboptions{sort&compress}

\usepackage[pdfstartview=FitH, colorlinks,
 linkcolor={red!60!black!},
 anchorcolor={green!80!black!},
 urlcolor={blue!80!black!},
 citecolor={blue!50!black!}]{hyperref}
\usepackage[table]{xcolor}

\journal{Physics Letters B}

\begin{document}


\begin{frontmatter}




\title{Double-magicity of proton drip-line nucleus $^{22}$Si with \textit{ab initio} calculation}


\author[ad1,ad2]{J.G. Li\corref{correspondence}}
\author[ad1,ad2]{H.H. Li}
\author[ad1]{S. Zhang}
\author[ad1,ad2]{Y.M. Xing \corref{correspondence}}
\author[ad1,ad2]{W. Zuo \corref{correspondence}}

\address[ad1]{CAS Key Laboratory of High Precision Nuclear Spectroscopy, Institute of Modern Physics,
Chinese Academy of Sciences, Lanzhou 730000, China}
\address[ad2]{School of Nuclear Science and Technology, University of Chinese Academy of Sciences, Beijing 100049, China}

\cortext[correspondence]{Corresponding author:\\ jianguo\_li@impcas.ac.cn (J.G. Li)\\
xym@impcas.ac.cn (Y.M. Xing) \\
zuowei@impcas.ac.cn (W. Zuo)}

\begin{abstract}

New magic numbers have been discovered in the neutron-rich region of the nuclear chart. However, there has been a lack of research on proton-rich nuclei. $^{22}$O, the mirror nucleus of $^{22}$Si,  is a double-magic nucleus bearing a high $E(2_1^+)$. Whether $^{22}$Si exhibits double-magic characters is an intriguing topic.
To investigate this matter, we utilized \textit{ab initio} valence space in-medium similarity renormalization group for $^{22}$Si/$^{22}$O, and their nearby nuclei. Our \textit{ab initio} calculations provide good descriptions for the double magicity of $^{22}$O, as well as the shell evolution of $N=14$ and $Z=14$ through $E(2_1^+)$. The computed $E(2_1^+)$ indicate that the closure of $Z=14$ sub-shell in proton-rich nuclei is weaker than the $N=14$ sub-shell closure in their mirror nuclei. Particularly, the calculated $E(2_1^+)$ of $^{22}$Si is 800 keV lower than the one of $^{22}$O. To further explore the magicity of $^{22}$Si, the mirror energy difference (MED) of $^{26}$Si/$^{26}$Mg,  $^{24}$Si/$^{24}$Ne, as well as $^{22}$Si/$^{22}$O are calculated. The results demonstrate that the calculated MEDs agree well with available experimental data, and the $E(2_1^+)$ values of $^{22,24,26}$Si are all lower than their respective mirror nuclei due to the Thomas-Ehrman shift with large $s_{1/2}$ occupation.
Moreover, our calculation provides that the many-body configurations of the low-lying state of $^{22}$Si/$^{22}$O are nearly identical despite the fact that the states bearing large MED. In conclusion, our \textit{ab initio} results suggest that  $^{22}$Si is a double magic nucleus, similar to its mirror nucleus $^{22}$O, albeit with a lower $E(2_1^+)$.

\end{abstract}


\end{frontmatter}

\section{Introduction} 
Shell structure is a characteristic feature of many-body fermionic systems, such as metallic clusters, atoms, and nuclei \cite{LEBOEUF2002127,RevModPhys.65.677,RevModPhys.92.015002}. Such structure is identified by the presence of magic numbers. Atomic nuclei consist of two types of interacting fermions, leading to a specific degree of freedom known as isospin \cite{Heisenberg1932berDB}. 
Experiments conducted over the last few decades have shown that the well-known magic numbers, such as 8,20,and 28, observed in stable nuclei disappear \cite{PhysRevC.41.1147,SORLIN2008602,RevModPhys.92.015002,PhysRevLett.96.032502,PhysRevLett.120.232501,PhysRevLett.105.102501} and new magic numbers emerge in neutron-rich nuclei far from stability line \cite{wienholtz2013masses,steppenbeck2013evidence,PhysRevC.92.034316,PhysRevLett.96.012501}. 
For instance, the magic numbers $N=8$ and $N=20$ vanish in the vicinity of $^{12}$Be \cite{PhysRevLett.96.032502} and $^{32}$Mg \cite{PhysRevLett.105.252501}, respectively, and the new magic number $N = 14$, 16, 32, and 34 appear in $^{22,24}$O \cite{PhysRevC.92.034316,PhysRevLett.96.012501} and $^{52,54}$Ca \cite{wienholtz2013masses,steppenbeck2013evidence}, respectively.

Isospin symmetry is a fundamental assumption in nuclear physics and has been  demonstrated to be highly accurate \cite{BENTLEY2007497,Heisenberg1932berDB, PhysRev.51.106}.
It was introduced by Heisenberg, who considered the proton and neutron as two distinct states of the same particle, differing only in the projection of isospin ($t_z$) \cite{Heisenberg1932berDB, PhysRev.51.106}.
Based on this assumption and irrespective of electromagnetic effects, \textit{i.e.}~with exact isospin symmetry, mirror nuclei with interchanged protons and neutrons should have identical energy levels. Additionally, the magic numbers of protons and neutrons should be same in principle, if charge independence holds. 
However, experiments show differences in excitation energy of analogue states in mirror nuclei that follow this pattern up to a few tens or hundreds of keVs \cite{PhysRevLett.89.142502,2006ARNPS..56..253M}. 
In particular, there exists a significant mirror energy difference (MED) in the states of mirror nuclei, where the valence protons of proton-rich nuclei occupy weakly-bound or unbound $s$ or $p$-wave, while the corresponding mirror neutron $s$ or $p$-wave is bound. This effect is referred to as the Thomas-Ehrman shift (TES) \cite{PhysRev.88.1109, PhysRev.81.412}. Although Coulomb effects are the primary cause of isospin-symmetry breaking (ISB), theoretical investigations indicate that the charge dependence of nuclear forces also plays a role therein. An intriguing question pertains to whether the robustness of neutron and proton shell closures are the same or not.

$^{22}$Si, the lightest nucleus with isospin projection of $T_Z = -3$, is of significant interest in nuclear physics. Its mirror nucleus, $^{22}$O, with proton and neutron shell closures of $Z= 8$ and $N= 14$, has been identified as a double magical nucleus with a high excitation energy of first $2_1^+$ state ($E(2_1^+)$) of roughly 3 MeV, which is higher than nearby nuclei \cite{THIROLF200016,PhysRevLett.96.012501}. 
The $\beta$-decays of $^{22}$O/$^{22}$Si to the first $1_1^+$ states of $^{22}$F/$^{22}$Al exhibit the largest mirror asymmetry of $\beta$ decay \cite{PhysRevLett.125.192503}. 
Moreover, numerous states with large MEDs have been observed in low-lying states of this region of nuclei (see Ref. \cite{PhysRevC.107.014302} and reference within). Theoretical calculations have also revealed that the TES effect occurs in low-lying states of $^{22}$Si \cite{PhysRevC.107.014302}. Consequently, the question of whether $^{22}$Si exhibits double magicity is an intriguing topic, and the well-developed valence space in-medium similarity renormalization group (VS-IMSRG) approach is employed in the present work to address this issue.

The structure of this paper is oganized as follows. First, the theoretical \textit{ab initio} VS-IMSRG approach employed in this paper is briefly introduced. The shell evolutions of shell closure at neutron number $N=14$ and proton number $Z=14$ are explored based on $E(2_1^+)$. Following that, spectra, MED values and average occupations of low-lying states in $^{26}$Si/$^{26}$Mg,  $^{24}$Si/$^{24}$Ne, and $^{22}$Si/$^{22}$O are presented, and the intriguing topic of whether $^{22}$Si exhibits double magicity is discussed. Finally, one proceeds to the summary of the paper.


\section{Method}
We start from an intrinsic A-body Hamiltonian, which is as follows:
\begin{equation}
 H=\sum_{i=1}^{A}\left(1-\frac{1}{A}\right) \frac{\boldsymbol{p}_{i}^{2}}{2 m}+\sum_{i<j}^{A}\left(v_{i j}^{\mathrm{NN}}-\frac{\boldsymbol{p}_{i} \cdot \boldsymbol{p}_{j}}{m A}\right)+\sum_{i<j<k}^{A} v_{i j k}^{3 \mathrm{N}},
 \label{H_intrinsic}
\end{equation}
where $\boldsymbol{p_i}$ is the nucleon momentum in the laboratory, and $m$ is mass of the nucleon. 
$v^{\rm NN}$ and  $v^{\rm 3N}$ are the two-body (\textit{NN}) and three-nucleon ($3N$) interactions, respectively.

The Hamiltonian of above Eq. (\ref{H_intrinsic}) can be rewritten with normal ordering with respect to reference state $|\Phi\rangle$ \cite{PhysRevLett.118.032502}:
\begin{equation}
 \begin{aligned}
 H= & E+\sum_{ij} f_{ij}: a_{i}^{\dagger} a_{j}:+\frac{1}{4} \sum_{ijkl} \Gamma_{ijkl}: a_{i}^{\dagger} a_{j}^{\dagger} a_l a_k :\\
  &+\frac{1}{36} \sum_{i j k l m n} W_{i j k l m n}: a_{i}^{\dagger} a_{j}^{\dagger} a_{k}^{\dagger} a_{n} a_{m} a_{l}:,
 \label{H_T2}
  \end{aligned}
\end{equation}
where the strings of creation and annihilation operators obey $\langle\Phi|: a_{i}^{\dagger} \cdots a_{j}:| \Phi\rangle=0$. Indeed, the normal-ordered zero-, one-, and two-body parts, denoted by $E$, $f_{ij}$, and $\Gamma_{ijkl}$, respectively, contain the main contributions of $v^{3N}$, allowing one to neglect the numerically expensive normal-ordered three-body part $W_{ijklmn}$ of the Hamiltonian \cite{HERGERT2016165,doi:10.1146/annurev-nucl-101917-021120}.

In the VS-IMSRG approach, the single-particle Hilbert space is divided into core, valence, and outside spaces. The chosen valence space of the Hamiltonian comprises the main degrees of freedom to produce properties of low-lying states for nuclei investigated. The primary objective of VS-IMSRG is to construct an effective Hamiltonian of the valence space, which is decoupled from the core (e.g. $^{16}$O) and outside single-particle orbitals. This procedure considers both the excitation out of the core and outside space. The decoupling can be achieved by solving the following flow equation
\begin{equation}
\frac{dH(s)}{ds}=[\eta(s),H(s)],
\label{FE}
\end{equation}
with the anti-Hermitian generator 
\begin{equation}
\eta(s)\equiv\frac{dU(s)}{ds}U^{\dagger}(s)=-\eta^{\dagger}(s),
\end{equation}
where $U(s)$ is the unitary transformation operator.

In this work, we employ the well-established $NN + 3N$ interaction provided by the 1.8/2.0 (EM) potential, which has been demonstrated to systematically reproduce ground-state energies up to $^{132}$Sn \cite{PhysRevLett.126.022501,PhysRevLett.120.152503,PhysRevC.105.014302}. For the 1.8/2.0 (EM) potential, the initial chiral next-to-next-to-next-to-leading order (N$^3$LO) \textit{NN} force \cite{PhysRevC.68.041001} is softened by a similarity renormalization group (SRG) evolution \cite{PhysRevC.75.061001} using $\lambda_{\rm SRG} = 1.8$ fm$^{-1}$, and a cutoff $\Lambda = 2.0$ fm$^{-1}$ is chosen for the corresponding next-to-next-to-leading order (N$^2$LO) $3N$ interaction. The short-range low-energy constants $c_D$ and $c_E$ are optimized to reproduce the triton bounding energy and $^{4}$He radius \cite{PhysRevC.83.031301}. Within chiral EFT framework, charge-symmetry and charge-independence breaking effects are well considered \cite{PhysRevC.72.044001,MACHLEIDT20111}. The Coulomb force is also included in Eq. (\ref{H_intrinsic}). In practical calculations, the harmonic-oscillator (HO) basis is used to define the model space. Specifically, we use 15 HO major shells (i.e.~$e=2n+l \leq e_{\mathrm{max}}=14$) with the frequency $\hbar \omega =16$ MeV, and additional truncation on the $3N$ matrix elements are limited as well to $e_{3\text{max}} = 2n_a+2n_b+2n_c+l_a+l_b+l_c\leq 14$.

In this work, we utilize VS-IMSRG with ensemble normal-ordering (ENO) \cite{PhysRevLett.118.032502,imsrg_code} to generate the valence-space Hamiltonian. The VS-IMSRG code of Ref. \cite{imsrg_code} is employed for that matter. We construct the effective Hamiltonian within the full $sd$-shell valence space above the doubly magic nucleus $^{16}$O. 
\textcolor{blue}{Within the decoupling process, all operators are truncated at two-body levels.
The truncation leads to the emergence of induced three-body forces.
Importantly, the impact of this induced three-body force on observables is small in VS-IMSRG calculations with ENO, especially concerning excited energies. Similar works have been done in Ref. \cite{PhysRevC.98.054320,PhysRevC.103.044318}.}
The obtained effective Hamiltonian can be exactly diagonalized using the shell-model code of Ref.~\cite{MICHEL2020106978} and used to investigate the physics of interest in the present work.


\section{Results}


\subsection{The shell evolution at $N=14$}

\begin{figure}[!htb]
\includegraphics[width=1.00\columnwidth]{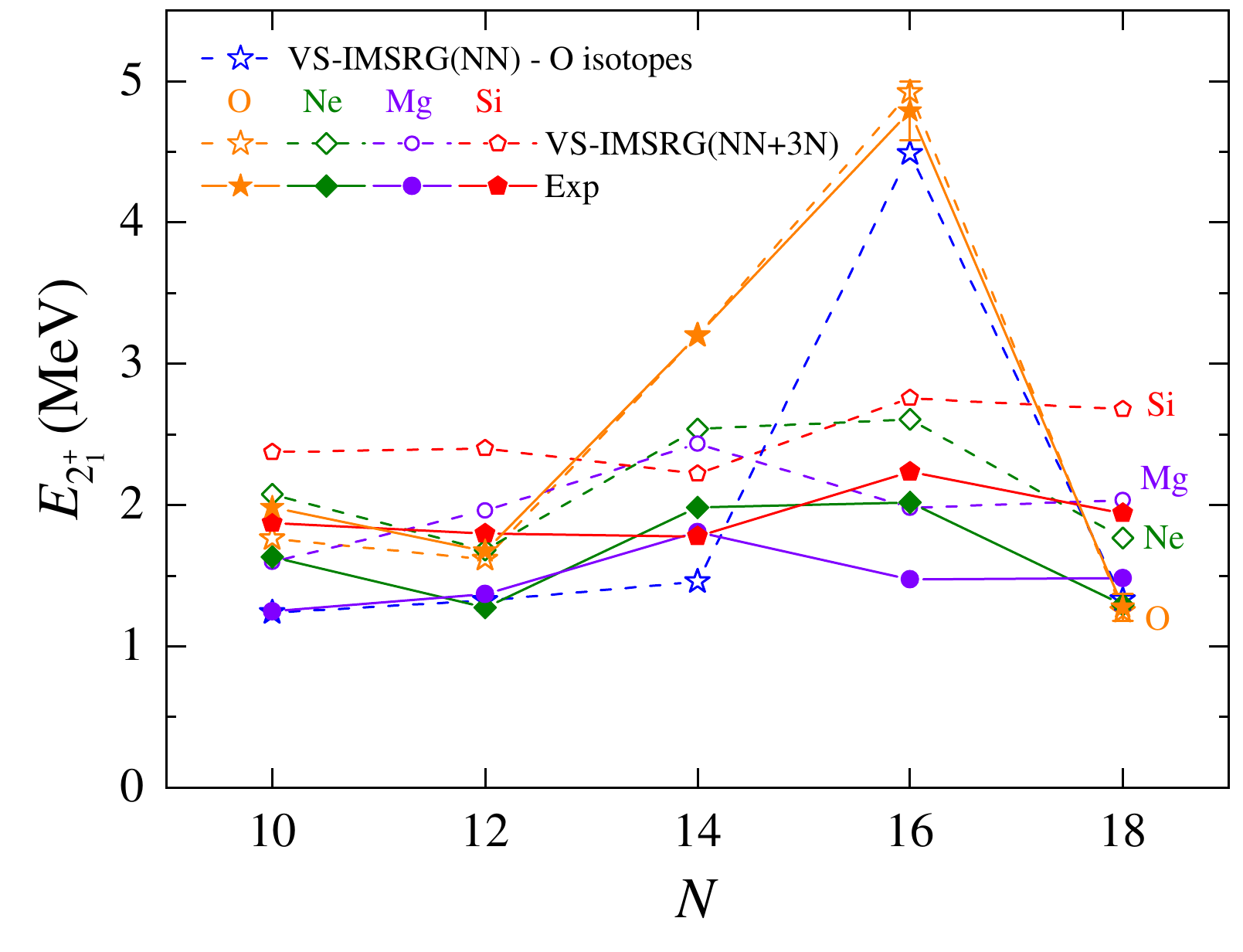}
\caption{The calculated excitation energy of first $2_1^+$ ($E(2_1^+)$) of oxygen, neon, magnesium and silicon isotopes with VS-IMSRG, along with experimental data \cite{ensdf}.}
\label{2+N}
\end{figure}

The $E(2_1^+)$  of even-even nuclei is the first observable that can provide information on the shell evolution. 
In the present work, we first investigate the shell evolution at neutron number $N=14$ in neutron-rich oxygen  isotopes using VS-IMSRG based on the chiral EM1.8/2.0 $NN +3N$ potential.
Figure \ref{2+N} presents the calculated $E(2_1^+)$ along with available experimental data \cite{ensdf}.
Our results show that the $E(2_1^+)$ of $^{22,24}$O are higher than those of nearby oxygen isotopes, implying that the $N=14$ and $N=16$ exhibit closed sub-shells and $^{22,24}$O are double magic nuclei. The results are in agreement with experimental data \cite{PhysRevC.92.034316,PhysRevLett.96.012501}. 
To make comparison, we also perform the \textit{ab initio} calculations using only chiral $NN$ force for oxygen isotopes, and the results are also shown in Fig.~\ref{2+N}. 
The $E(2_1^+)$  of neutron-rich oxygen isotopes obtained from the chiral $NN$  calculations are lower than those from the chiral $NN+3N$ calculations and experimental data \cite{ensdf,PhysRevC.92.034316,PhysRevLett.96.012501}, particularly in the case of $^{22}$O. The results reveal that the double magicity of $^{22}$O can not be reproduced without including the $3N$ force. 

To investigate the shell evolution at $N=14$ in neutron-rich neon, magnesium, and silicon isotopes, we calculated the $E(2_1^+)$, which are presented in Fig. \ref{2+N} as well.
As shown, although the calculated $E(2_1^+)$ of neon, magnesium and silicon isotopes are generally slightly higher, approximately 500 keV, than the experimental data, the trend of their evolution is in good agreement with experimental data.
Both the theoretical calculations and experimental data show that the $E(2_1^+)$  of $^{24,26}$Ne and $^{26}$Mg are higher than those of their nearby isotopes, and the $E(2_1^+)$ values of $^{22-30}$Si are close.
The results suggest that the $N=14$ sub-shell remains to exist in the neon and magnesium chains but disappears in silicon chain. Furthermore, as valence protons are added above the oxygen isotopes, the closure of $N=14$ sub-shell weakens, as evidenced by the decreasing trend of $E(2_1^+)$ from $^{22}$O to $^{28}$Si. Additionally, the $N=16$ sub-shell observed in the oxygen and neon isotopes also disappears in the magnesium and silicon isotopes.

\subsection{The shell evolution at $Z=14$}

\begin{figure}[!htb]
\includegraphics[width=1.00\columnwidth]{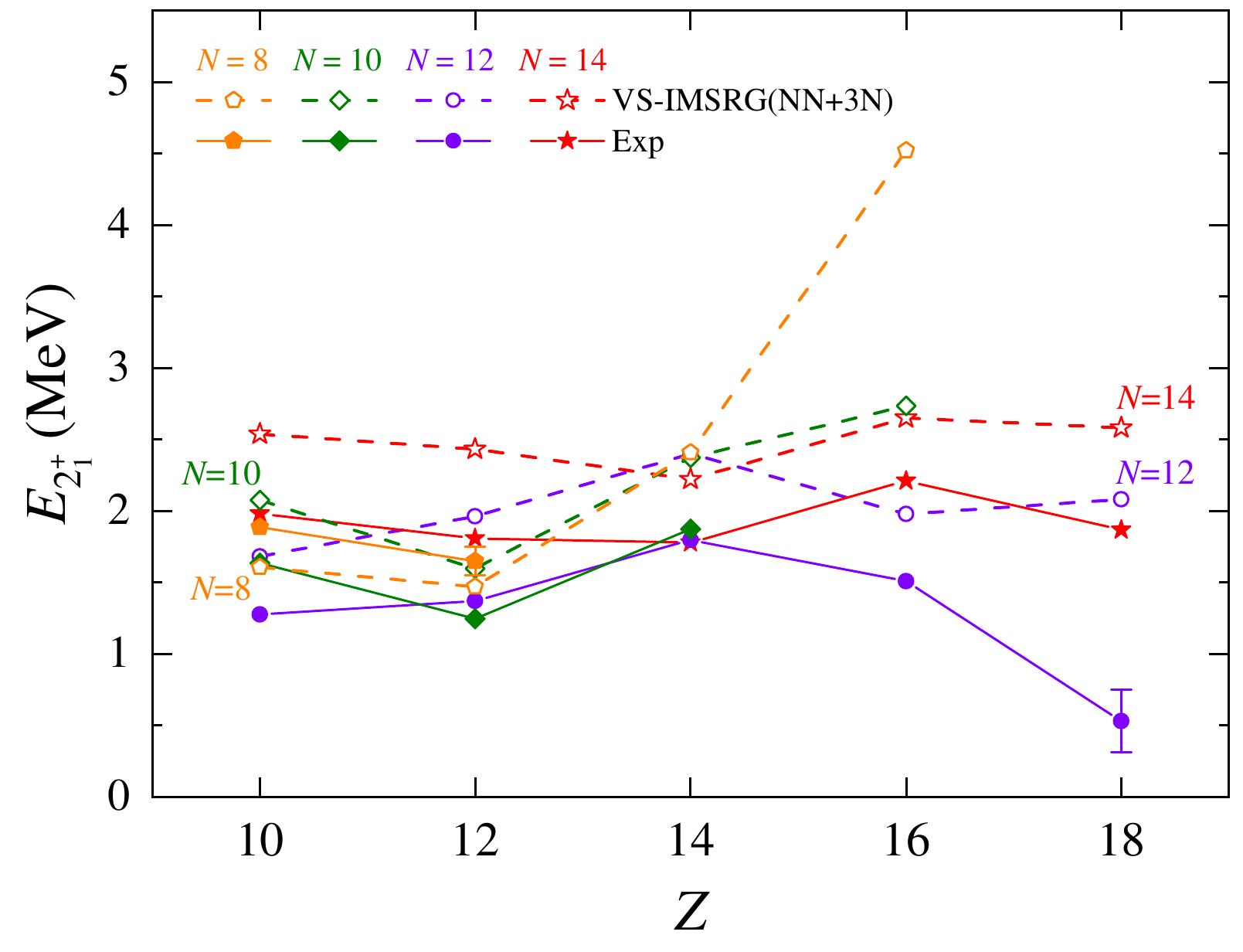}
\caption{Similar to Fig. \ref{2+N}, but for the $E(2_1^+)$ of $N =$ 8, 10, 12, and 14 isotones.}{}
\label{2+Z}
\end{figure}

Similar calculations have also been performed to analyze the closure of $Z=14$ sub-shell nearby Si isotopes. 
Our VS-IMSRG results for the $E(2_1^+)$ of $N=8$, 10, 12 and 14 isotones are presented in Fig. \ref{2+Z}, and compared with available experimental data \cite{ensdf}.
In contrast to the neutron-rich isotopes near $N=14$, there is limited experimental information about the excited states of proton-rich nuclei in the vicinity of the Si isotopes. 
Notably, proton drip-line nucleus $^{22}$Si, which is the mirror nucleus of $^{22}$O, currently lacks experimental data on its mass and spectra.
Nevertheless, our \textit{ab initio} VS-IMSRG calculations well reproduce the trend of  $E(2_1^+)$ evolution in $N=8$, 10, 12 and 14 isotones. 
Both the theoretical and experimental $E(2_1^+)$ suggest that the sub-shells of $Z=14$ and $N=14$ both disappear in $^{28}$Si.
Similar to the $N=14$ sub-shell shown in Fig. \ref{2+N}, the strength of $Z=14$ sub-shell is enhanced with the removal of neutrons from $^{28}$Si, and the magicity of $Z=14$ sub-shell present in $^{24}$Si, where the $E(2_1^+)$ of $^{24,26}$Si are higher than that of nearby nuclei $^{22,24}$Mg.

Our \textit{ab initio} VS-IMSRG calculation predicts that the $E(2_1^+)$ of $^{22}$Si is about 2.4  MeV which is consistent with both realistic and phenomenological Gamow shell model calculations \cite{ZHANG2022136958,PhysRevC.100.064303}, and many-body perturbation theory based on $NN+3N$ \cite{PhysRevLett.110.022502}.
 As shown in Fig.~\ref{2+Z}, this calculated $E(2_1^+)$ of $^{22}$Si is slightly higher than that of adjacent nucleus $^{20}$Mg but is comparable to that of $^{24,26}$Si isotopes. This is in contrast to the obviously enhanced strength of $N=14$ closure sub-shell when moving from $^{26}$Mg to $^{22}$O, as shown in Fig.~\ref{2+N}.
Whether $^{22}$Si is a double magic nucleus is difficult to see from the calculated $E(2_1^+)$. 


\begin{figure*}[!htb]
\includegraphics[width=2.00\columnwidth]{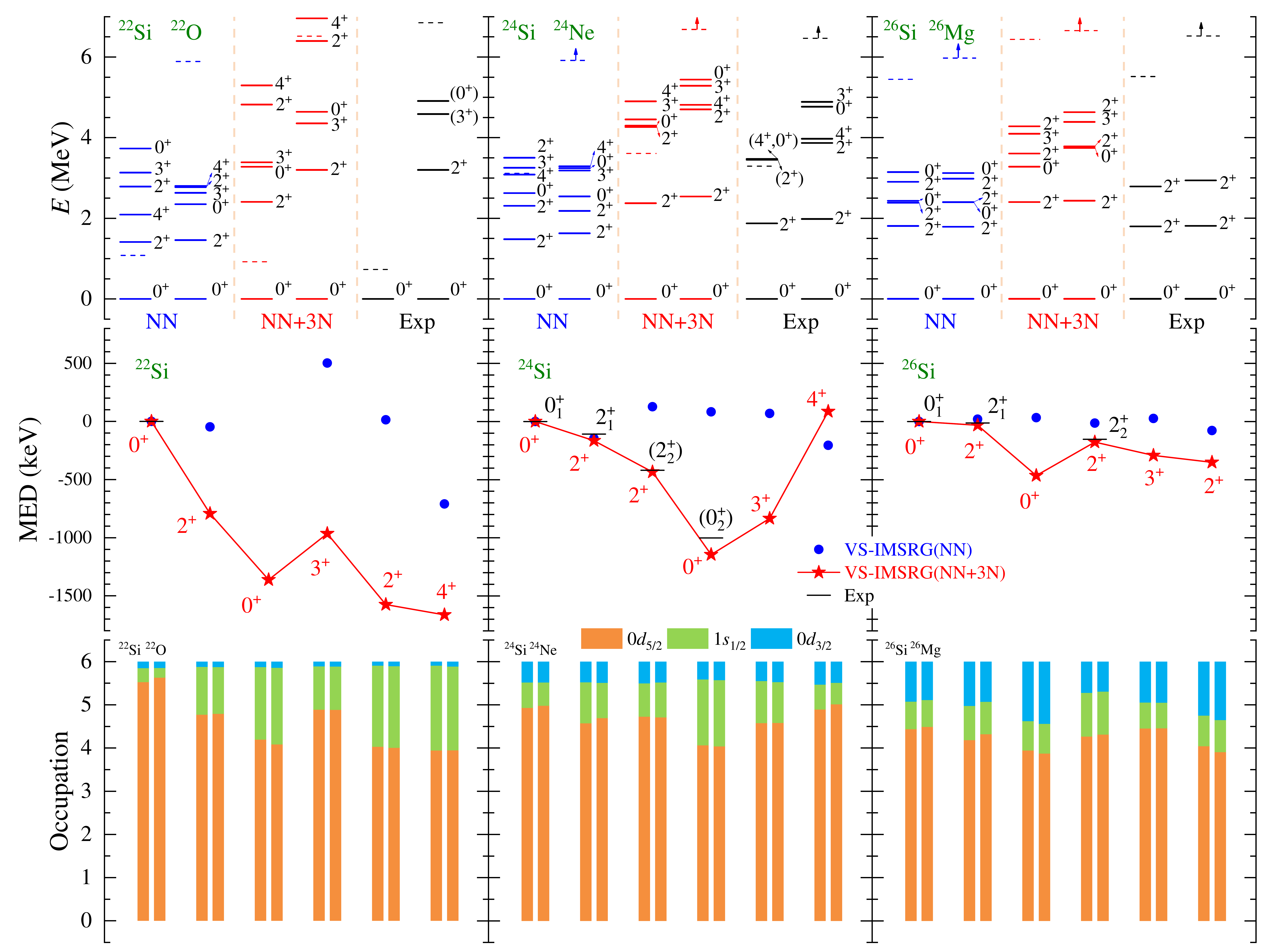}
\caption{\textit{Ab initio} VS-IMSRG calculations of spectra, MED, and average occupations of mirror nuclei $^{22}$Si/$^{22}$O,  $^{24}$Si/$^{24}$Ne, and $^{26}$Si/$^{26}$Mg with only $NN$ and $NN + 3N$ interactions, along with experimental data \cite{ensdf,PhysRevC.101.031303}. Shown average occupations correspond to the valence protons and neutrons in proton- and neutron-rich nuclei, respectively.}{}
\label{Mirror-Si}
\end{figure*}

\subsection{The mirror energy difference of proton-rich Si dripline isotopes}
The above \textit{ab initio} calculations reveal that $^{22}$O is a double magic nucleus, while its mirror nucleus $^{22}$Si has a lower $E(2_1^+)$ value compared to that of $^{22}$O.
To further understand the situation, we investigated the isospin-symmetry breaking in mirror energy difference between even-even $^{22-26}$Si and their respective mirror nuclei using \textit{ab initio} VS-IMSRG with only $NN$ and $NN + 3N$ interaction. The calculated low-lying spectra and MEDs for mirror states in $^{22}$Si/$^{22}$O, $^{24}$Si/$^{24}$Ne, and $^{26}$Si/$^{26}$Mg are presented in Fig. \ref{Mirror-Si}, together with  available experimental data \cite{ensdf,PhysRevC.101.031303}.
We can see that the calculated low-lying spectra and MEDs of mirror nuclei $^{24}$Si/$^{24}$Ne and $^{26}$Si/$^{26}$Mg utilizing $NN + 3N$ interaction are in good agreement with experimental data, particularly the MEDs of low-lying states.
It is worth noting that the \textit{ab initio} calculations with only $NN$ force give compressed spectra of low-lying states for $^{24}$Si/$^{24}$Ne and $^{26}$Si/$^{26}$Mg. Furthermore, the calculated MEDs values with only $NN$ force are almost zero and are smaller than both the experimental data and theoretical calculations with $NN+3N$ force, showing significant discrepancies  with the experimental data \cite{ensdf,PhysRevC.101.031303}.

To comprehensively investigate the isospin symmetry breaking, the average occupations of the low-lying states are calculated, and are depicted in Fig. \ref{Mirror-Si}.
The presented average occupations correspond to valence protons and neutrons in proton-rich and neutron-rich nuclei respectively. 
According to the calculated average occupations, we can see that the occupations of valence protons and neutrons in mirror states are virtually identical.
A state with a large MED indicates that it has significant isospin symmetry breaking, and possesses higher occupations of $1s_{1/2}$ orbital compared to that of ground state, in which the  $1s_{1/2}$ orbital is bound in neutron-rich nuclei, while  weakly-bound or unbound in proton-rich nuclei.
The phenomenon, known as TES effect, is particular evident in $0_2^+$ state of $^{24}$Si/$^{24}$Ne, where the MED of the mirror state is about 1 MeV, and the $1s_{1/2}$ orbital occupations is approximately 1.5, as opposed to the ground states, where the occupation of the $1s_{1/2}$ orbital is approximately 0.5. 
Additionally, the large occupations of $0d_{3/2}$ orbital also contribute to MED. Detail discussions regarding the mechanism of the states with large MED values can be found in our previous work \cite{PhysRevC.107.014302}.
The good agreement of the calculations of  $^{24}$Si/$^{24}$Ne and $^{26}$Si/$^{26}$Mg provides impetus for further exploration into the properties of mirror nuclei $^{22}$Si/$^{22}$O, which are of particular interest in the present work.

Our \textit{Ab initio} calculation with $NN+3N$ provide a good description of the low-lying states of $^{22}$O, and predict significant isospin symmetry breaking  with large MED values in the low-lying states of mirror nuclei $^{22}$Si/$^{22}$O. However, calculations with only $NN$ potential give that the low-lying spectra of $^{22}$Si/$^{22}$O are excessively compressed in comparison to experimental data and $NN+3N$ calculations, and fail to reproduce the double magicity of $^{22}$O.
Additionally, the many-body wave functions obtained from $NN$ calculations for $^{22}$Si/$^{22}$O give that the ground states are dominated by $(0d_{5/2})^4(1s_{1/2})^2$ configuration, while the $(0d_{5/2})^6$ configuration dominantes in the $0_2^+$ excited states. These results are inconsistent with $NN+3N$ calculations and phenomenological shell model calculation \cite{PhysRevC.101.064312}.

Our calculations predict that the $E(2_1^+)$ of $^{22}$Si and $^{22}$O are about 2.4 and 3.2 MeV, respectively.
The corresponding large MED value of about $-800$ keV indicates a significant isospin symmetry breaking.
The $NN+3N$ calculations reveal that the $2_1^+$ states of $^{22}$Si/$^{22}$O have large occupations of $1s_{1/2}$ orbital, which is unbound in $^{22}$Si but deeply-bound in  $^{22}$O.
Due to the unbound $1s_{1/2}$ in proton drip line nucleus $^{22}$Si, which exhibits strong coupling with higher $s_{1/2}$ orbitals, the effective single-particle energy gap between $0d_{5/2}$ and $1s_{1/2}$ orbitals in $^{22}$Si is smaller than that in $^{22}$O. 
As a result, the $E(2_1^+)$ of $^{22}$Si is smaller than that of $^{22}$O.
Interestingly, despite the significant difference in the gaps between $0d_{5/2}$ and $1s_{1/2}$ in $^{22}$Si/$^{22}$O, the calculated average occupations of these nuclei are almost identical.
Moreover, the $\pi/\nu(0d_{5/2})^6$ configuration dominates the ground states of $^{22}$Si/$^{22}$O by about 78\% and 83\%, respectively.
Therefore, our calculations suggest that the doubly-magic properties of $^{22}$Si persist, although with a small $E(2_1^+)$ due to TES effect. 
Moreover, our \textit{ab initio} calculations using $NN+3N$ interaction also provide predictions for the excitation energies and MED of higher excited states in $^{22}$Si/$^{22}$O.

\textcolor{blue}{ Continuum coupling plays an important role in describing the properties of dripline nuclei \cite{0954-3899-36-1-013101,PhysRevLett.108.242501,PhysRevLett.109.032502,PhysRevC.99.061302,physics3040062,ZHANG2022136958}.
To have physical weakly bound and unbound many-body wave functions extended in coordinate space, it would be preferable to perform calculations in the Berggren basis \cite{BERGGREN1968265}, where bound, resonance, and continuum are treated on the same footing. 
However, the inclusion of the continuum leading to the proliferation of states aggravates the intruder-state problem, making VS-IMSRG calculations within the Berggren basis challenging \cite{doi:10.1146/annurev-nucl-101917-021120}.
In the VS-IMSRG calculations, we preferred to perform the calculation using a large number of HO shells. $N_{\rm max} = 14$ is employed in the present calculations, which has been seen to partially describe the extended asymptotes of weakly bound and unbound many-body states.
This is corroborated by the fact that the calculated MEDs of $sd$-shell nuclei using VS-IMSRG align well with experimental results \cite{PhysRevC.107.014302}.}

\begin{figure}[!htb]
\includegraphics[width=1.00\columnwidth]{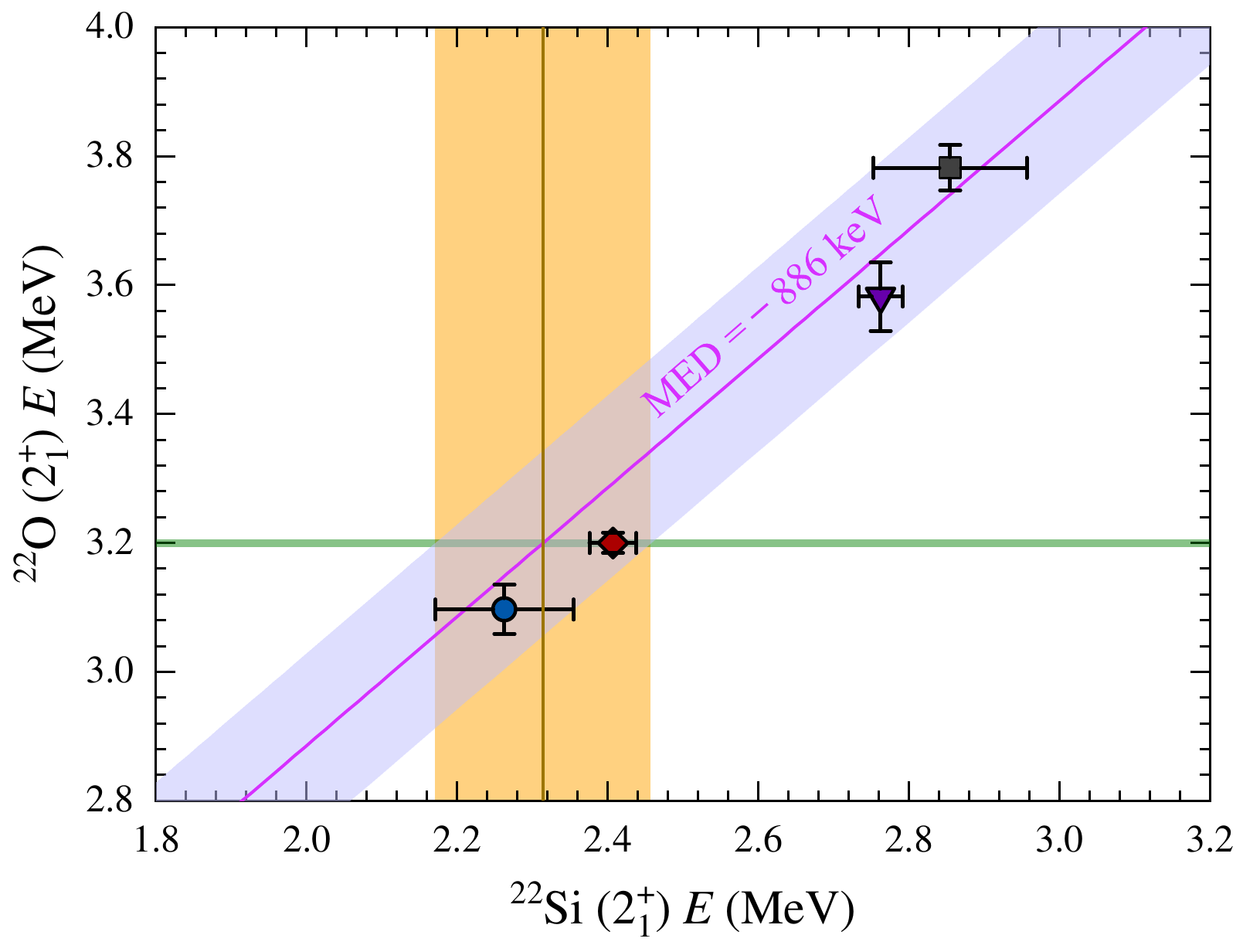}
\caption{Correlation between the energies of the $2_1^+$ excited state in $^{22}$O and $^{22}$Si mirror partner, obtained from the interactions NNLO$_{\rm sat}$(gray square), 1.8/2.0 (EM) (red diamond), NN+3N(lnl) (purple triangle), and NNLO$_{\rm opt}$(blue circle). The error bars estimate uncertainties from enlarging the model space from $N =12$ to $N =14$. The thin horizontal line marks the known energy of the $2_1^+$ state in $^{22}$O.
The MED values of the $2_1^+$ are estimated to be approximately $-886$ keV based on the calculated.}{}
\label{22Si-22O}
\end{figure}

\textcolor{red}{Additionally, VS-IMSRG calculations using NNLO$_{\rm sat}$ \cite{PhysRevC.91.051301}, NNLO$_{\rm opt}$ \cite{PhysRevLett.110.192502} and NN+3N(lnl) \cite{PhysRevC.101.014318} interaction have also been performed. 
For the NN+3N(lnl) interaction, the induced three-body force is neglected by employing a large similarity renormalization group cutoff of $\lambda = 2.6$ fm$^{-1}$.
Notably, the NNLO$_{\rm sat}$ has been shown to provide accurate descriptions for nuclear radii \cite{PhysRevC.91.051301}.
The results for $E(2_1^+)$ for $^{22}$Si and $^{22}$O are presented in Fig. \ref{22Si-22O}.
We employ the methodology outlined in Ref.~\cite{PhysRevLett.117.172501} for uncertainty estimation.
The error bars on the individual data points are determined from the model-space truncation from $N = 12$ to $N = 14$, and they remain smaller than 100 keV for all interactions used.
While the obtained $E(2_1^+)$ for $^{22}$O and $^{22}$Si show slight variations based on the employed interaction, a robust correlation is evident in the calculated $E(2_1^+)$ for $^{22}$Si and $^{22}$O, underpinned by a MED value of approximately $-$886 keV.
The distinct horizontal line in Fig. \ref{22Si-22O} denotes the known $E(2_1^+)$ state in $^{22}$O, and its intersection with the diagonal band gives our theoretical estimate 2.18 MeV $\precsim E(2_1^+) \precsim$ 2.42 MeV for $^{22}$Si. 
Notably, the results from the NN+3N(lnl) and 1.8/2.0 EM interactions fall within the crossed uncertainty band. 
Consequently, the consistent result furthers the credibility of our predictions for the double magicity of $^{22}$Si.}

\section{summary}
We have utilized the \textit{ab initio} in-medium similarity renormalization group to investigate the shell evolution at $Z=14$ and $N=14$, as well as the double magicity of $^{22}$O/$^{22}$Si based on the calculated $E(2_1^+)$. 
Our calculations indicate that $^{22}$O exhibits the double magicity, while the sub-shell closure of $N=14$ gradually weakens as the isotopes change from oxygen to silicon, and the results are consistent with experimental data.
In addition, the calculated $E(2_1^+)$ remains almost the same in $^{22.24,26,28}$Si,  which is in contrast to the mirror case.
To analyze the contradiction, we have calculated the mirror energy difference of $^{22,24,26}$Si with their respective mirror nuclei.
Our \textit{ab initio} calculations with $NN+3N$ interaction well reproduce the mirror energy differences of low-lying states of $^{24}$Si/$^{24}$Ne and $^{26}$Si/$^{26}$Mg. 
More importantly, our calculation predict that the $E(2_1^+)$ of $^{22}$Si is lower about 800 keV than that of its mirror nucleus $^{22}$O, which is mainly caused by the Thomas-Ehrman shift effect with large $s_{1/2}$ occupation.
However, the calculated many-body configurations of ground states of $^{22}$Si/$^{22}$O are, nonetheless, fairly similar.
In conclusion, our \textit{ab initio} calculations suggest that the $^{22}$Si is a double magic nucleus, similar to its mirror nucleus $^{22}$O, despite bearing a smaller $E(2_1^+)$ than that of $^{22}$O.

\section{Acknowledgments}

 This work has been supported by the National Natural Science Foundation of China under Grant Nos.  12205340, 11975282 and 11905261;  the Gansu Natural Science Foundation under Grant No. 22JR5RA123 and 23JRRA614; 
 the Youth Innovation Promotion Association of Chinese Academy of Sciences under Grant No. 2021419; the Strategic Priority Research Program of Chinese Academy of Sciences under Grant No. XDB34000000; the Key Research Program of the Chinese Academy of Sciences under Grant No. XDPB15; the State Key Laboratory of Nuclear Physics and Technology, Peking University under Grant No. NPT2020KFY13. 
 This research was made possible by using the computing resources of Gansu Advanced Computing Center.



\section*{References}

\bibliographystyle{elsarticle-num_noURL}
\bibliography{Ref}





\end{document}